\documentclass[aps,amsfonts,pra,twocolumn]{revtex4}

\pdfoutput=1

\usepackage{epsfig,amsmath,amssymb,bm,epsf,graphicx,psfrag}
\usepackage[all]{xy}
\usepackage{graphicx}
\usepackage{color}

\newcommand{\ncd}{\newcommand}
\ncd{\RMbra}{\langle {\cal{R}}(r,m)|}
\ncd{\RMket}{|{\cal{R}}(r,m)\rangle}

\newtheorem{Theorem}{Theorem}

\newtheorem{Lemma}{Lemma} 

\newtheorem{Def}{Definition}

\begin{document}

\title{Contextuality in Measurement-based Quantum Computation}

\author{Robert Raussendorf}
\affiliation{Department of Physics and Astronomy, University of British
Columbia, Vancouver, British Columbia V6T 1Z1, Canada}
\date{\today}

\begin{abstract}
 We show, under natural assumptions for qubit systems, that measurement-based quantum computations (MBQCs) which compute a non-linear Boolean function with high probability are contextual. The class of contextual MBQCs includes an example which is of practical interest and has a super-polynomial speedup over the best known classical algorithm, namely the quantum algorithm that solves the `Discrete Log' problem.
\end{abstract}

\maketitle

\section{Introduction}

While numerous quantum algorithms have been found that offer polynomial or super-polynomial speedups over their classical counterparts \cite{DL,Grov,QW}, the precise quantum mechanical origin of this speedup remains unknown. The prominent candidates---entanglement \cite{E}, superposition and interference \cite{SI}, largeness of Hilbert space \cite{CI}---provide an intuitive understanding in many situations. Yet, as a whole, the phenomenology so far uncovered does not lend itself to a simple interpretation \cite{Stab} - \cite{TETBU}.  

Here we turn our attention to a different characterization of non-classicality, namely contextuality \cite{KS, Bell}, and study its relation to computational power. We choose measurement-based quantum computation (MBQC) \cite{RB01} as our setting. The starting point for this investigation is the observation by Anders and Browne \cite{AB} that one of Mermin's proofs \cite{Merm} of the Kochen-Specker theorem \cite{KS} can be converted into a simple measurement-based quantum computation. We are led to ask whether the connection between MBQC and contextuality exhibited by this example is accidental, or whether it holds in general. The main finding of this paper is that, under quite natural assumptions for multi-qubit systems,  all MBQCs which compute a non-linear Boolean function with sufficiently high success probability are contextual. 

For MBQC, the separation between linear and non-linear functions is fundamental. Every MBQC requires a classical control computer for adjusting measurement bases according to  the computational input and for converting measurement outcomes into computational output. This classical side-processing is all linear. Evaluating non-linear functions is out of reach for such a classical control computer without access to quantum resources.

This paper is organized as follows. In Section~\ref{Set}, we review Anders and Browne's example, and define the setting of MBQC and notions of contextuality we will use. In Section~\ref{CnCP} we present three results on the interplay between contextuality and the non-linearity of the computational output, Theorems~\ref{KSCT}, \ref{ProbExt} and \ref{ProbExtC}. We point out that the class of contextual MBQCs contains a computation which is of actual algorithmic interest, i.e., achieves a super-polynomial speedup over the best-known classical algorithm. It is the MBQC-variant of the quantum algorithm for the `Discrete Log' problem \cite{DL,Mosca}. We conclude with a discussion in Section~\ref{Disc}.

\section{The setting}
\label{Set}

Measurement-based quantum computation (MBQC) \cite{RB01} is a model of universal quantum computation in which a quantum algorithm is implemented solely by local measurements on a fixed initial state. The choice of measurement bases determines the algorithm to be implemented, and correlations among the measurement outcomes reveal the result of the computation. The computational power of this scheme is fully determined by the initial quantum state. For suitable initial states such as cluster states, MBQC is universal.

\subsection{Computation and contextuality: a first example}
\label{GHZ}

To discuss the connection between measurement-based quantum computation (MBQC) and contextuality it is instructive to first look at an example. Following Anders and Browne \cite{AB}, we consider a three-party Greenberger-Horne-Zeilinger (GHZ) \cite{GHZ} state
$
|\text{GHZ}\rangle = \frac{|000\rangle+|111\rangle}{\sqrt{2}}
$
which can be used to execute a deterministic OR-gate within the framework of MBQC. 

While standard electronic devices routinely perform OR-gates without quantum-mechanical spooky action, this result offers a structural insight into MBQC. Namely, it is known that every MBQC requires a classical control computer that converts the classical input into measurement settings, and the measurement outcomes into computational output. This classical control computer is capable of doing only one type of operation--- addition mod 2. It is thus not classically universal, and indeed very limited. Now, having access to GHZ-states and local projective measurements promotes this control computer to classical universality. Thus, in the described setting, the access to quantum resources vastly increases the set of computable functions. 

What is more, Anders and Browne's construction repurposes an existing proof \cite{Merm} of the Kochen-Specker theorem \cite{KS}  into a quantum mechanical computation. 

The computation takes two bits of input, $i_1$ and $i_2$, and outputs a single bit $o\equiv i_1 \vee i_2$. It proceeds as follows. Step 1: the settings for the local measurements on the three qubits are calculated from the input $i_1$, $i_2$. For either of the three qubits, a priori the observables $O_k = X_k,Y_k$ can be measured (here and in the following, $\sigma_x\equiv X,\;\sigma_y\equiv Y,\;\sigma_z\equiv Z$), and we use the binary variable $q_k$ to encode the choice. If $q_k=0 \,(1)$  then $X_k$ ($Y_k$) is measured. The measurement setting $\textbf{q}=(q_1,q_2,q_3)$ is related to the input $\textbf{i} = (i_1,i_2)$ via $q_1=i_1$, $q_2=i_2$, $q_3 = i_1 + i_2 \mod 2$.  Step 2: The observables $O_k(q_k)$ are being measured whereby the measurement outcomes $s_k \in \{0,1\}$ are obtained. Here, if the measured eigenvalue of the Pauli observable $O_k$ was $+1$ ($-1$) then $s_k=0$ ($s_k=1$). Step 3: The parity $o \equiv s_1+s_2+s_3 \mod 2$ of the three measurement outcomes is computed and outputted.

It is easily verified that this procedure does indeed compute the desired OR-gate. First note that $|\mbox{GHZ}\rangle$ is an eigenstate with eigenvalue 1 of the following operators: 
\begin{equation}
  \label{GHZcorr}
  X_1X_2X_3,\;-X_1Y_2Y_3,\;-Y_1X_2Y_3,\; -Y_1Y_2X_3. 
\end{equation}
The outcomes of the local $X$-and $Y$-measurements selected by the input $\textbf{i}$ in the above procedure are thus strictly correlated or anti-correlated. Specifically, if $i_1=i_2=0$ then the measured observables are $X_1,X_2,X_3$. The measurement outcomes $s_1,s_2,s_3$ are individually random, but because of $X_1X_2X_3|\mbox{GHZ}\rangle = |\mbox{GHZ}\rangle$, $o(0,0) = s_1 + s_2 + s_3 \mod 2 =0$ with certainty. Likewise, if $i_1=0,i_2=1$ then the measured observables are $X_1,Y_2,Y_3$. As before, the local measurement outcomes $s_1,s_2,s_3$ are individually random, but because of the relation $X_1Y_2Y_3|\mbox{GHZ}\rangle = - |\mbox{GHZ}\rangle$ we find that $o(0,1)=s_1+s_2+s+3 \mod 2 =1$ with certainty. The remaining two cases are analogous, and we thus verify the logical table of the OR-gate.\medskip

The present implementation of the OR-gate is quantum mechanical, and one may ask whether contextuality of quantum mechanics is brought to bear in this process. Let's try to construct a non-contextual  hidden variable model for the  `pre-determined' measurement outcomes $x_1,x_2,x_3,\, y_1,y_2,y_3 \in \mathbb{Z}_2$ of the observables $X_1, X_2, X_3, Y_1, Y_2, Y_3$ potentially measured in the realization of the gate. Since the measured observables are all local, such a hidden variable model is also local (a special case of being non-contextual). Going through each entry in the logical table of the OR-gate, the following relations are imposed
\begin{equation}
\label{MermKS}
\begin{array}{rcl}
  x_1 + x_2 + x_3 \mod 2 &=& 0,\\
  x_1 + y_2 + y_3 \mod 2 &=& 1,\\
  y_1 + x_2 + y_3 \mod 2 &=& 1,\\
  y_2 + y_2 + y_3 \mod 2 &=& 1.
\end{array}
\end{equation}
Adding up these four equations, we find $2(x_1+y_1+x_2+y_2+x_3+y_3)\mod 2 = 0 =1$. Contradiction!---No assignment of pre-determined local measurement outcomes reproduces the correlations required for an OR-gate in the present three-party setting. The argument we have just stated is, in fact, Mermin's state-dependent proof of the Kochen-Specker theorem in dimension 8 \cite{Merm}. We find that a proof of contextuality of quantum mechanics can be re-purposed as a (simple) quantum computation.\medskip

We may take this example a step further and consider the following modifications: (i) Flipping  some observables $O_k(q_k) \longrightarrow -O_k(q_k)$. (ii) Using instead of $|\mbox{GHZ}\rangle$ some other state from the GHZ-family, i.e. a simultaneous eigenstate of the observables in Eq.~(\ref{GHZcorr}), but with eigenvalues $-1$ for some of them. 

Since both the changes can be implemented by local unitary operation, one expects contextuality to remain unaffected. And indeed, while the right hand side of Eq.~(\ref{MermKS}) does change under such transformations, the number of entries 1 always remains odd. Hence, the contradiction for non-contextual hidden variable models persists. As for the computed Boolean function, it also changes but always remains non-linear. We are thus led to ask: Is there a link between the non-linearity of a Boolean function computed in MBQC and non-contextuality of such a computation? This is the question which we will subsequently investigate. To do so, we first need to define our precise setting of MBQC and notion of contextuality.

\subsection{The general setting of measurement-based computation}

The above example using a GHZ-state may serve as a first illustration of MBQC, but it misses two aspects: (i) MBQC is universal for quantum computation, and (ii) the measurements in MBQCs can be  temporally ordered. The latter is a consequence of the randomness inherent in quantum measurement. To prevent this randomness from creeping into the logical processing, measurement bases need to be adjusted to measurement outcomes already obtained. This leads to a partial temporal order of the measurement events; See Fig.~\ref{LinExpl}.

One may consider an MBQC-scenario with $n$ parties,  $k$ measurement settings at each party, and $l$ possible outcomes for each of those measurements. However, for the confines of this paper we restrict our attention to the case of two measurement settings per party and two outcomes for each local measurement, i.e. $k=l=2$. This is a natural choice when the local quantum systems are qubits. We further impose a restriction on the classical side-processing in MBQC which is required for the adjustment of measurement bases according to previously obtained outcomes and for obtaining the computational output from the local measurement outcomes. Namely, all such processing should be mod 2-linear. 

The relations between contextuality and computational power described in Section~\ref{CnCP}  (or at least their present proofs) crucially depend on this linearity. Therefore, before making definitions, we need to motivate such linear constraints. In this regard, we note that in the GHZ-example of the previous section all classical side-processing is indeed mod 2 linear. However, the main justification for imposing mod 2 linear relations of  classical side processing is that they are sufficient for quantum-mechanically universal MBQC on cluster states \cite{RB01}.  The origin of linearity in the classical side-processing is explained in Appendix~\ref{LCPR}.  We note that MBQC schemes with different classical processing exist \cite{Eisert}.

We now introduce the notion of {\em{l2-MBQC}} for ``MBQC with mod 2 linear classical processing''.

\begin{Def}[l2-MBQC]\label{l2MBQC}An l2-MBQC is a mea\-sure\-ment-based quantum computation with classical input $\textbf{i}$ and classical output $\textbf{o}$, where the measurements driving the computation are all local and satisfy the following properties:
\begin{enumerate}
\item{\label{C1}For each party $k$, $k=1..n$, there is a binary choice for the measurement basis, $q_k \in \{0,1\}$.}  
\item{\label{C2}For each party $k$ and each $q_k \in \{0,1\}$, the measurement outcome is binary-valued, $s_k \in \{0,1\}$.} 
\item{\label{C3}The computational output  $\textbf{o}$ is bitwise a parity of measurement outcomes $\textbf{s} = (s_1,s_2,..,s_n)^T$,
\begin{equation}
\label{outCorr}
\textbf{o} = Z \textbf{s} \mod 2.
\end{equation}}
\item{\label{C4}The choice of measurement bases $\textbf{q} = (q_1,q_2,É,q_n)^T$ is related to the measurement outcomes $\textbf{s}$ and the binary-valued classical input $\textbf{i} = (i_1,.., i_l)^T$ via
\begin{equation}
\label{MB}
\textbf{q} = T\textbf{s} + Q \textbf{i} \mod 2.
\end{equation}}
\item{\label{C5}For a suitable ordering of the parties $1 .. n$, the matrix $T$ in Eq.~(\ref{MB}) is lower triangular with vanishing diagonal.}
\end{enumerate}
\end{Def}
The reason for imposing the Condition~\ref{C5} is that if and only if $Q$ is   strictly lower triangular  with respect to a suitable ordering of parties, the given MBQC is {\em{runnable}}, i.e., measurement bases depend only on measurement outcomes that have already been obtained.  

\subsection{Contextuality}
\label{ctxt}

In a hidden variable model (HVM), in stark contrast to quantum mechanics, measurement outcomes exist prior to measurement, and are merely ``revealed''.  Non-contextuality means the following: Let an observable $A$ be measured jointly with one of the compatible observables $B$ or $C$, and $B$  is incompatible with $C$. An HVM is non-contextual if the `pre-existing' measurement outcome $\lambda(A)$ for  $A$ is independent of whether $A$ is measured jointly with $B$ or  with $C$. Non-contextual HVMs cannot reproduce all predictions of quantum mechanics in Hilbert spaces of dimension $\geq 3$. This is the content of the Kochen-Specker theorem~\cite{KS}.

Non-contextuality is not compromised by the classical communication required in l2-MBQC. In quantum mechanics, two observables are compatible if and only if the corresponding Hermitian operators commute. Consider two parties, $a$ and $b$, with to-be-measured observables $O_a(q_a)$ and $O_b(q_b)$ where $q_a,q_b \in \{0,1\}$. The values of $q_a$ and $q_b$ are specified by prior measurement outcomes; see Eq.~(\ref{MB}). Independent of the values of $q_a$ and $q_b$, the observables $O_a$ and $O_b$ always commute because they are local to different tensor product factors of the underlying Hilbert space. \medskip

\begin{figure}
\begin{center}
\includegraphics[width=0.8\columnwidth]{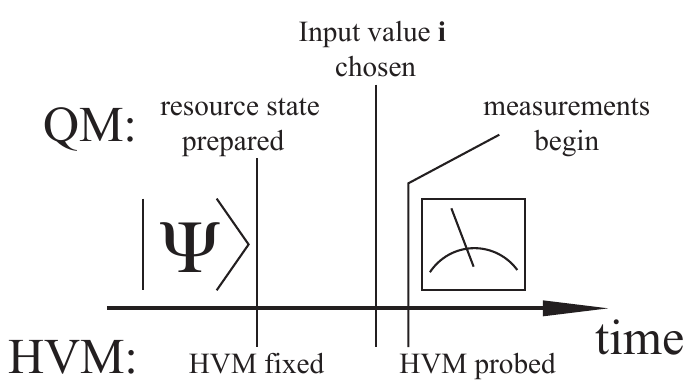}
\end{center}
\caption{\label{LOCvsCONT}Timeline of events in an MBQC. 1. The resource quantum state is prepared. 2) The computational input $\textbf{i}$ is chosen. 3) The measurements comprising the computation are performed, and their outcomes are processed. The hidden variable model attempting to describe the computation is fixed {\em{before}} the input $\textbf{i}$ is chosen.}
\end{figure}

We follow the sheaf-theoretic notion of contextuality developed by Abramsky and Brandenburger  \cite{Abram1}. Below we restate  from \cite{Abram1} the definitions of the notions  required for the present discussion, namely `section', `measurement context', `phenomenological model', `strongly contextual' and `contextual'. We adapt to the setting of l2-MBQC where suitable; for example, in our discussion measurement outcomes will always be in $\mathbb{Z}_2$. Our notion of `measurement context' accommodates measured observables depending on previously obtained measurement outcomes.

{\em{Sections.}} Be $X$ the set of measurements and $\mathbb{Z}_2$ the set of outcomes for each individual measurement. For all $U\subseteq X$, a section over $U$ is a function $s: U \rightarrow \mathbb{Z}_2$. It describes measurement outcomes $s=(s(O_1),s(O_2),..,s(O_n))$, $O_i \in U$. A section over $X$ is called `global'.  We denote the set of sections $s$ over $U$ by $\mathbb{Z}_2^U$, and by ${\cal{E}}$ the map ${\cal{E}}: U \mapsto \mathbb{Z}_2^U$. Contextuality is about the non-existence of global sections.\smallskip

{\em{Measurement contexts.}}  A measurement context is a set $C \subset X$ of compatible measurements. The set ${\cal{M}}$ of measurement contexts has the following two properties: (i) $X=\bigcup_{C\in {\cal{M}}}C$, and (ii) For $C,C' \in {\cal{M}}$, if $C \subseteq C'$ then $C=C'$. The second property says that contexts are maximal sets of compatible measurements.

A feature of measurement contexts in l2-MBQC that is not explicitly addressed in \cite{Abram1} is the adaptivity of local measurement settings according to previously obtained measurement outcomes. In l2-MBQC, the measurement at party $k$ depends on the $m$-bit input $\textbf{i}$ and the measurement outcomes $\left.s\right|_{{\cal{P}}(k)}$ obtained in the past ${\cal{P}}(k)$ of $k$; c.f. Eq.~(\ref{MB}). 

Contexts are labeled by the basis choice $\textbf{q}$, 
\begin{equation}
C(\textbf{q}) =\{O_k(q_k),\, k = 1..n \}.
\end{equation}
The measurement record $\textbf{s}$ appearing in Eqs.~(\ref{outCorr}),(\ref{MB}) and the section $s$ are related via 
\begin{equation}
\textbf{s}=\left. s\right|_{C(\textbf{q})}.
\end{equation}
Since $\textbf{q}$ depends on $\textbf{s}$ via Eq.~(\ref{MB}), this looks like self-consistency condition. Given $s$ and $\textbf{q}$, it is a priory not clear whether $\textbf{s}$ exists, and if it it does, whether it is unique. However, due to the runnability condition on the matrix $Q$ in Eq.~(\ref{MB}), a unique solution $\textbf{s}$ does indeed always exist. The set $\{1,..,n\}$ of qubits is divided into smaller sets ${\cal{Q}}_0$, ${\cal{Q}}_1$, .. , ${\cal{Q}}_{\text{last}}$ which are measured one after the other. The measurement bases for any given set only require knowledge of the measurement outcomes from prior sets. Choosing the bases for the measurements in ${\cal{Q}}_0$ requires no knowledge of measurement outcomes, and the set ${\cal{Q}}_0$ is therefore measured first. Using the measurement outcomes from ${\cal{Q}}_0$, $\left. \textbf{q}\right|_{{\cal{Q}}_1}$ can  be obtained and ${\cal{Q}}_1$ be measured, and so on. Correspondingly, the components of $\textbf{s}$ can be extracted one set ${\cal{Q}}_i$ at a time.\smallskip

{\em{A phenomenological model}} $e$ is a set of probability distributions $\{e_C|\,C \in {\cal{M}}\}$ such that for each measurement context $C$ and measurement outcome $\textbf{s} \in {\cal{E}}(C)$, $e_C(\textbf{s})$ is the probability of obtaining $\textbf{s}$ within $C$. We may consider any process that begins with a preparation and ends with a measurement---quantum, classical or other---as a phenomenological model.\smallskip

{\em{Strong contextuality.}} We define the set ${\cal{S}}_e$ of global sections that only predict possible events,
\begin{equation}
  \label{setSe}
  S_e:=\left\{\textbf{s} \in {\cal{E}}(X) |\, e_C(\textbf{s}|_C)>0, \forall C \in {\cal{M}}\right\}.
\end{equation}
\begin{Def}[Strong contextuality]\label{strongcont}
A phe\-no\-me\-no\-lo\-gi\-cal model $e$ is strongly contextual if $S_e=\emptyset$.
\end{Def}
A model $e$ with $S_e = \emptyset$ is definitely contextual since no assignment $\textbf{s}\in {\cal{E}}(X)$ of `pre-determined' measurement outcomes---and no probability distribution over such assignments---can reproduce it. Such contextuality is called ``strong'' because it implies other forms of contextuality \cite{Abram1}.\medskip

{\em{Contextuality.}} Consider a phenomenological model $e$ for which the set $S_e$ is not necessarily empty. Even if ${\cal{S}}_e \neq \emptyset$, the model $e$ may still  be contextual. While consistent assignments for the pre-determined measurement outcomes exist, no probability distribution over those assignments may reproduce the probability distributions in $e$. 

We label the elements in $S_e$ by a hidden variable $\lambda$, i.e., $S_e= \{\textbf{s}(\lambda),\lambda \in \Lambda\}$. Each $\textbf{s}(\lambda)$ induces a set of probability distributions $\{d_C(\lambda), C \in {\cal{M}}\}$ over the measurement records in all measurement contexts. Then, a probability distribution $q$ of the hidden variable $\lambda$ induces a set of probability distributions $\{d_C(q)=\sum_{\lambda \in \Lambda}q(\lambda)\, d_C(\lambda), C \in {\cal{M}}\}$ over the measurement records in all measurement contexts. 

\begin{Def}[Contextuality]\label{ProbCont}
A phe\-no\-menological model $e$ is contextual if the set of linear equations
$$
  e_C=\sum_{\lambda \in \Lambda}q(\lambda)\, d_C(\lambda),\; \forall C \in {\cal{M}},
$$
has no solution $q$ with $q(\lambda)\geq 0$ for all $\lambda \in \Lambda$. If it has such a solution, the model is non-contextual.
\end{Def}
Contextuality is weaker than strong contextuality: the latter always implies the former. The converse does not hold. The Bell scenario is contextual but not strongly contextual \cite{Abram1}. {\em{Remark:}} In \cite{Abram1}, `contextual' is called `probabilistically non-extendable'.\smallskip
 
Now consider a phenomenological model $e$ with a non-contextual part $K$ and a general no-signalling part $Q$, 
\begin{equation}
\label{convdec}
e =p\,K+(1-p)Q,\;\; 0\leq p \leq 1.
\end{equation}
That is, for all contexts $C \in {\cal{M}}$ and all sections $\textbf{s} \in {\cal{E}}(C)$ we have $e_C(\textbf{s}) =p\,K_C(\textbf{s})+(1-p)Q_C(\textbf{s})$.  We call the supremum of $p$ over all convex decompositions Eq.~(\ref{convdec}) the non-contextual fraction of $e$. A phenomenological model $e$ is called {\em{maximally contextual}} if its non-contextual fraction is zero. We then have the relation \cite{Abram1}
\begin{Theorem}\label{StrongMax}
A model is strongly contextual if and only if it is maximally contextual.
\end{Theorem} 
This concludes our review of the required notions of contextuality from \cite{Abram1}.\medskip

{\em{Remark:}} MBQC \cite{RB01} uses only local observables to drive the computation. An alternative approach therefore is  to relate MBQC to non-locality of quantum mechanics rather than contextuality. If so, a complication is posed by the adaptive choice of measurement bases in MBQC. The local parties exchange classical messages in order to adjust measurement, which runs counter to the assumption of locality. However, post-selection may be employed to restore locality. For the kind of post-selection required, Bell-type inequalities can still be derived which indicate non-locality of MBQCs \cite{HB11}.

\section{Contextuality and com\-pu\-ta\-tio\-nal power}
\label{CnCP}

In this section we present our results on the interplay between contextuality and non-linearity in MBQC.

\subsection{The deterministic case}

We call an l2-MBQC {\em{deterministic}} if it computes the vector-valued Boolean function $\textbf{o}$ with unit probability for every allowed input $\textbf{i} \in \mathbb{Z}_2^m$. We regard l2-MBQCs as phenomenological models. That is, an l2-MBQC $M$ is a set of probability distributions, $M = \{M_{C(\textbf{q})} |\, C(\textbf{q}) \in {\cal{M}}\}$.

\begin{Theorem}
\label{KSCT}
Be M an l2-MBQC which deterministically evaluates a vector $\textbf{o}$ of  Boolean functions on an  input $\textbf{i}$.  If $\textbf{o}(\textbf{i})$ is non-linear mod 2 in $\textbf{i}$ then M is strongly contextual.  
\end{Theorem}

{\em{Proof of Theorem~\ref{KSCT}.}}  We show that if $M$ is not strongly contextual, then $\textbf{o}(\textbf{i})$ is mod-2-linear in $\textbf{i}$. The result then follows by negation.

If $M$ is not strongly contextual then there exists a non-contextual HVM consistently assigning values to the observables in the set $X = \bigcup_{\textbf{q}}C(\textbf{q})$, i.e., $S_M$ is non-empty.  For each party $k = 1.. n$, there are at most two possible measurement bases, labeled by $q_k=0$ and $q_k=1$, respectively (Property~\ref{C1} in Def.~\ref{l2MBQC}). Therefore, $X \subset \tilde{X}: = \{O_k(q_k=0), O_k(q_k=1), \, \forall  k=1..n\}$, but $X$ may be strictly smaller than $\tilde{X}$. First, consider the case where both $O_k(q_k=0)$ and $O_k(q_k=1)$ are in $X$.  In the non-contextual HVM, these observables have pre-existing  outcomes  $s_i(q_k=0)$ and $s_i(q_k=1)$ that are independent of the context $C(\textbf{q})$. Since  the measurement outcomes $s_k$ are binary (Property~\ref{C2}), and any function defined on only two points is linear, these outcomes can be expressed in terms of two binary variables $c_k$ and $d_k$,
\begin{equation}
  \label{setA2}
  s_k(q_k) \equiv c_k \oplus d_k q_k.
\end{equation}
The pair $[s_k(q_k=0),s_k(q_k=1)]$ and the pair $[c_k,d_k]$ contain the exact same information. Keep in mind that  $c_k$ and $d_k$ depend upon the chosen $s \in {\cal{S}}_M$.

Next, consider the case where, for a given party $k$, only one of the two values of $q_k$ occurs for all $\textbf{s} \in {\cal{S}}_M$ and all $\textbf{i} \in \mathbb{Z}_2^m$. This can happen only if the $k$th row of $Q$ is identically zero. Then, Eq.~(\ref{setA2}) does still hold. The only difference is that  $c_k$, $d_k$ are no longer unique. 

Thus, the relation Eq.~(\ref{setA2}) holds for all parties $k=1..n$. We convert it into vector form,
$
	\textbf{s}(\textbf{q}) \equiv \textbf{c} \oplus D \textbf{q},
$
where $D = \text{diag}(d_k)$. Inserting Eq.~(\ref{MB}) (see Property~\ref{C4} of Def.~\ref{l2MBQC}) into this relation, we obtain
$$
(I \oplus D\,T)\textbf{s} \equiv \textbf{c}  \oplus D\,Q \, \textbf{i}.
$$ 
Therein, the matrix $I \oplus D\,T$ is invertible because of Property~\ref{C5}. We can thus always solve for $\textbf{s}$, and, using Eq.~(\ref{outCorr}) (see Property~\ref{C3} in Def.~\ref{l2MBQC}), we obtain for the classical output $\textbf{o}$ of the computation
\begin{equation}
\label{ClOut}
 \textbf{o} = \textbf{c}' + Q'\, \textbf{i} \mod 2,\end{equation}
where $\textbf{c}'=Z(I\oplus DT)^{-1}\textbf{c}$ and $Q'=Z(I\oplus DT)^{-1}DQ$. We emphasize that $\textbf{c}'$ and $Q'$ may depend on the choice $s \in {\cal{S}}_M$ via $\textbf{c}$ and $D$, c.f. Eq.~(\ref{setA2}).  Therefore, $\textbf{o}(\textbf{i})$ is linear in $\textbf{i}$ given a particular $s \in{\cal{S}}_M$. To make explicit the potential dependence of $\textbf{o}$ on the choice of the global section, we choose a reference section $s_0 \in {\cal{S}}_M$; hence $ \textbf{o} (s_0)= \textbf{c}'(s_0) + Q'(s_0)\, \textbf{i} \mod 2$. Now, for any  choice $s_\lambda \in {\cal{S}}_M$, for all $\textbf{i} \in \mathbb{Z}_2^m$ we must have
$$
\textbf{o} (s_0)= \textbf{c}'(s_0) \oplus Q'(s_0)\textbf{i} = \textbf{c}'(s_\lambda) \oplus Q'(s_\lambda)\textbf{i} = 
\textbf{o} (s_\lambda),
$$
for otherwise the computation would not be deterministic. Thus, for all $s_\lambda \in {\cal{S}}_M$, $\textbf{c}'(s_\lambda)=\textbf{c}'(s_0)=:\textbf{c}'$ and $Q'(s_\lambda)=Q'(s_0)=:Q'$. The output $\textbf{o}(\textbf{i})$ in Eq.~(\ref{ClOut}) is thus linear mod 2 in $\textbf{i}$.
$\Box$\medskip

It should be noted that besides the simple initial example of \cite{AB} (c.f. Section~\ref{GHZ}), the class of contextual l2-MBQCs  contains a quantum algorithm with super-polynomial speedup over the best known classical counterpart. Namely, `Discrete Log' can be made deterministic in the circuit model \cite{Mosca}, and its translation an l2-MBQC thus falls under Theorem~\ref{KSCT}. 

\subsection{The probabilistic case}
\label{Prob_Ext}

One may be interested in probabilistic extensions of Theorem~\ref{KSCT} for at least two reasons: (i) Quantum-mechanical phenomena are in general statistical rather than deterministic, and (ii)  The algebraic underpinning of the deterministic setting just described is unstable to even infinitesimal perturbations, calling into question any possible relation to experiments \cite{DM}.
 
To begin, we need a notion of probabilistic function evaluation. 
\begin{Def}\label{ProbFu}A  procedure $\tau$ probabilistically evaluates a vector-valued Boolean function $\textbf{o}(\textbf{i})$ on an $m$-bit input $\textbf{i} \in V^m$ with success probability $p_S$ if 
$$
\min_{\textbf{i} \in V^m} \text{Prob}\left(\tau(\textbf{i}) = \textbf{o}(\textbf{i})\right) = p_S.
$$
\end{Def}
The realization of probabilistic function evaluation as an l2-MBQC has the phenomenological model
\begin{equation}
  \label{tauDec1}
  M = p K + (1-p) Q,
\end{equation} 
where $Q$ is a contextual non-signalling model and $K$ a non-contextual model, such that the non-contextual fraction $p$ is maximized (subject to the constraint $0 \leq p \leq 1$). The model $M$ is contextual for all $p<1$ and, with Theorem~\ref{StrongMax}, strongly contextual for $p=0$. Now, consider the case where $M$ deterministically evaluates a non-linear Boolean function, $p_S=1$. By Theorem~\ref{KSCT}, $M$ then is strongly contextual. Hence, by Theorem~\ref{StrongMax}, $M$ is maximally contextual, $p=0$. We now ask: For probabilistic evaluation of a Boolean function with an l2-MBQC $M$, how much can the success probability $p_S$ drop for the computation to remain contextual ($p<1$)? This question leads us to
\begin{Theorem}
\label{ProbExt}
Be $M_{p_s}$ an l2-MBQC that probabilistically evaluates a vector of non-linear Boolean functions on $m$ bits of input, with success probability $p_S$. If $p_S>1-\frac{1}{2^m}$ then $M_{p_S}$ is contextual.
\end{Theorem}

We thus find that contextuality persists within a finite interval around the point of strong contextuality. However, we also find that for general non-linear functions, the contextuality threshold for $p_S$ approaches unity exponentially fast in the input size $m$. This result can be significantly improved for special non-linear Boolean functions, as will be discussed below.

In preparation for the proof of Theorem~\ref{ProbExt}, we define the distance $\nu$ of a Boolean function $o$ to the closest linear Boolean function, $\nu = \min_{l \in \text{lin.B.f}}\text{wt}(o \oplus l)$. For a vector $\textbf{o}$ of Boolean functions, we define the distance to the closest linear function as $\nu:= \min_{\textbf{l} \in \text{lin.B.f}} \sum_{\textbf{i} \in V^m}  \delta(\textbf{o}(\textbf{i}) \oplus \textbf{l}(\textbf{i}))$. We then have
\begin{Lemma}
\label{ProbExtLem}
Be $M_{p_s}$ an l2-MBQC that evaluates with success probability $p_S$ a vector-valued Boolean function on $m$ input bits with distance $\nu$ to the closest linear function. If $p_S>1-\frac{\nu}{2^m}$ then $M_{p_S}$ is contextual.
\end{Lemma}

{\em{Proof of Lemma~\ref{ProbExtLem}.}} We decompose the non-contextual part $pK$ in the r.h.s. of Eq.~(\ref{tauDec1}) as
$$
pK = \sum_k q_k L_k,
$$ 
where all $L_k$ are non-contextual models corresponding to l2-MBQCs which deterministically evaluate functions $\textbf{l}_k$, and all $q_k\geq 0$. By Theorem~\ref{KSCT}, all functions $\textbf{l}_k$ are linear mod 2.    We define ${\cal{L}}_k:=\{\textbf{i}\in V^m|\, l_k(\textbf{i}) \neq \textbf{o}(\textbf{i}) \}$. Then,
\begin{equation}
  \label{nucond}
  \nu \leq |{\cal{L}}_k|,\;\; \forall k.
\end{equation}
For any given input $\textbf{i} \in V^m$, the non-contextual part $pL$ in Eq.~(\ref{tauDec1}) contributes a portion $p_{\text{fail},L}(\textbf{i})$ to the probability of failure to output $\textbf{o}(\textbf{i})$, 
$p_{\text{fail},L}(\textbf{i}) = \sum_{k| \textbf{i} \in {\cal{L}}_k}q_k
$. The contextual part $(1-p)Q$ in Eq.~(\ref{tauDec1}) may also contribute, and the probability $p_{\text{fail}}(\textbf{i})$ of failure to output $\textbf{o}(\textbf{i})$ is thus the same or bigger. Summing over all $\textbf{i} \in V^m$, we have
$$
\sum_{\textbf{i}} p_{\text{fail}}(\textbf{i}) \geq \sum_{\textbf{i}}\sum_{k| \textbf{i} \in {\cal{L}}_k}q_k = \sum_k q_k |{\cal{L}}_k| \geq p\nu.
$$
Further, $\displaystyle{1-p_S = \max_{\textbf{i} \in V^m} p_{\text{fail}}(\textbf{i})}$, by Def.~\ref{ProbFu}. Also, the failure probability averaged over the $2^m$ input values $\textbf{i} \in V^m$ is smaller or equal to the maximal failure probability, $\displaystyle{\max_{\textbf{i} \in V^m} p_{\text{fail}}(\textbf{i}) \geq 2^{-m}  \sum_{\textbf{i} \in V^m} p_{\text{fail}}(\textbf{i})}.$
Combining the last three relations, we find
$$
2^m(1-p_S)\geq p\nu.
$$
To maintain contextuality, $p$ must be bounded away from unity. With the last relation, this is guaranteed if $p_S > 1 - \frac{\nu}{2^m}$. $\Box$\medskip

{\em{The Proof of Theorem~\ref{ProbExt}}} is now straightforward. For any non-linear Boolean function, $\nu \geq 1$. Theorem~\ref{ProbExt} now follows from Lemma~\ref{ProbExtLem} with the choice $\nu=1$. $\Box$\medskip

If we consider all non-linear Boolean functions, no $\nu$ larger than 1 can be chosen for Theorem~\ref{ProbExt}, since  $o=\prod_{k=1}^m i_k$ has $\nu=1$. However, if we restrict to special functions then the range of $p_S$ for which the computation remains contextual can be significantly extended. In this regard, we recall from MacWilliams and Sloane \cite{MWS} the following 
\begin{Def}
A Boolean function $f(v_1,...,v_m)$, for $m$ even, is called `bent' if the Hadamard transform coefficients $\hat{F}(\textbf{u})$ given by
$
\hat{F}(\textbf{u}) = \sum_{\textbf{v} \in V^m} (-1)^{\textbf{u}\cdot\textbf{v}+f(\textbf{v})}
$
are all $\pm 2^{m/2}$.
\end{Def}
For bent functions we note \cite{MWS}
\begin{Theorem}\label{mwsBF} A bent function $f(v_1,.., v_m)$ is further away from any linear function
$$
  a_0+ \sum_{i=1}^ma_iv_i
$$ 
than any other Boolean function. More precisely, $f(v_1,..,v_m)$ is bent iff the corresponding vector $\textbf{f}$ has a distance $2^{m-1}\pm 2^{m/2-1}$ from every codeword of the Reed-Muller code ${\cal{R}}(1,m)$. If $f$ is not bent, then $\textbf{f}$ has a distance less than $2^{m-1}-2^{m/2-1}$ from some codeword of ${\cal{R}}(1,m)$.
\end{Theorem}
Using this result, we obtain
\begin{Theorem}\label{ProbExtC}
Be $M_{p_S}$ an l2-MBQC that evaluates with success probability $p_S$ a bent function on an even number $m$ of bits. Then $M_{p_S}$ is contextual if $p_S>\frac{1}{2}+\left(\frac{1}{2}\right)^{m/2+1}$.
\end{Theorem}

{\em{Proof of Theorem~\ref{ProbExtC}.}} With Theorem~\ref{mwsBF}, we can choose $\nu=2^{m-1}-2^{m/2-1}$, and the result follows directly from Lemma~\ref{ProbExtLem}. $\Box$\medskip

The low threshold of $p_S$ in Theorem~\ref{ProbExtC} is worth of note. Consider the special case of a single output bit (which for any l2-MBQC can be obtained by discarding the other output qubits). Then, for large values of $m$, the output of $M_{p_s}$ can be very close to completely random, and yet $M_{p_S}$ remains contextual.

\section{Discussion}
\label{Disc}

In summary, we have shown that l2-MBQCs cannot be described by non-contextual hidden variable models if they compute non-linear Boolean functions with sufficiently high probability of success. The probability threshold depends on the Boolean function in question. It is very close to 1 for products of high degree but close to 1/2 for bent functions. In addition, we would like to draw attention to the following two points:\smallskip

\paragraph{Beyond the quantum case.} Although we have stated Theorems~\ref{KSCT}, \ref{ProbExt}, \ref{ProbExtC} for measurement-based {\em{quantum}}  computations, they hold in more general scenarios than quantum theory. An example for such more general (and hypothetical) scenarios are Popescu-Rohrlich boxes, which maximally violate the CHSH inequality. Neither the definitions \cite{Abram1} of contextuality applied here nor the proofs of Lemma~\ref{ProbExtLem} and Theorems~\ref{KSCT}, \ref{ProbExt}, \ref{ProbExtC} use properties of quantum mechanics. Required are the binary choice for the measurement basis and a binary measurement outcome for each party, the linear processing relations Eqs.~(\ref{outCorr}), (\ref{MB}), and runnability (Property~\ref{C5} in Def.~\ref{l2MBQC}). \smallskip

\paragraph{Comparison: Hard sampling without contextuality.} In \cite{Hob} it is shown that MBQC with local measurements in fixed bases on a separable resource state can produce probability distributions of computational output which are classically hard to sample from, unless the polynomial hierarchy in complexity theory collapses at its third level. The measurements and resource state are both classical; yet these MBQCs achieve something that is believed to be classically hard. How is this possible?

Despite appearances, there is non-classicality in the MBQCs of \cite{Hob}: the resource state, while being separable, is hard to create classically (it can efficiently be created by quantum-mechanical means). The present description of MBQC as a phenomenological model does not take this kind of non-classicality into account.

The question that now arises is whether such computational hardness of sampling is at odds with Theorems~\ref{KSCT} and \ref{ProbExt} of this paper. It seems so: Theorems~\ref{KSCT} and \ref{ProbExt} state than an l2-MBQC becomes contextual as soon as it achieves anything beyond outputting parities. Theorem~2 of \cite{Hob} states that MBQCs can efficiently perform a classically hard task without being contextual. However, there is no contradiction. The functions probabilistically evaluated by the MBQCs in \cite{Hob} are all mod 2 linear, as required by Theorems~\ref{KSCT} and \ref{ProbExt}. But that doesn't render the output trivial. While in the present paper the goal is to compute a function with high probability of success, i.e., to output peaked probability distributions, in \cite{Hob} the probability distributions don't have to have a dominant peak at all. The only, and rather different, requirement on them is that they are hard sample from. For the MBQCs discussed here, contextuality is revealed only after multiple executions with different values of input. In the case of \cite{Hob}, all hardness is contained in the output distribution for a single input value.\medskip

\textbf{Acknowledgements.} The author thanks Pradeep Sarvepalli and Tzu-Chieh Wei for discussions. This work is funded by NSERC, Cifar, PIMS and IARPA.

\appendix

\section{Linear classical processing relations in MBQC}
\label{LCPR}

\begin{figure}[h]
\begin{center}
\begin{tabular}{ll}
(a) \\
&\includegraphics[width=0.7\columnwidth]{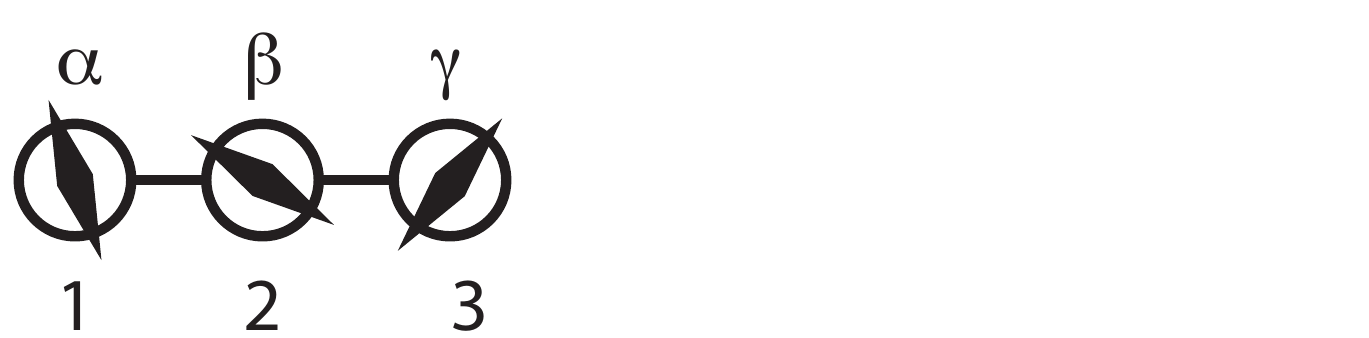} \vspace{2mm}\\
(b) \\
&\includegraphics[width=0.7\columnwidth]{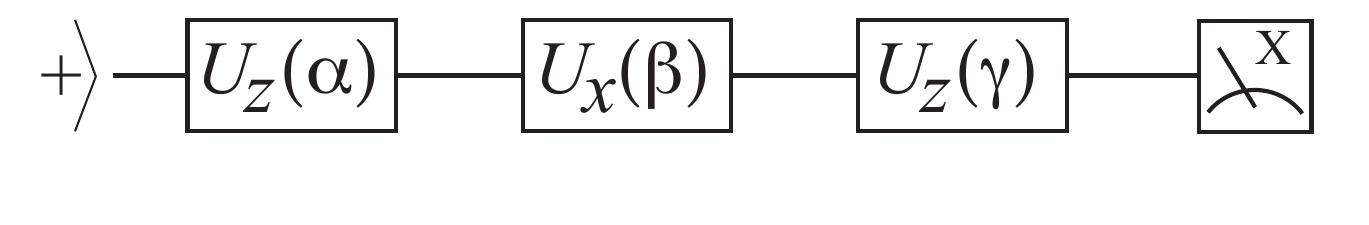} \vspace{2mm}\\
(c) \\
&\includegraphics[width=0.7\columnwidth]{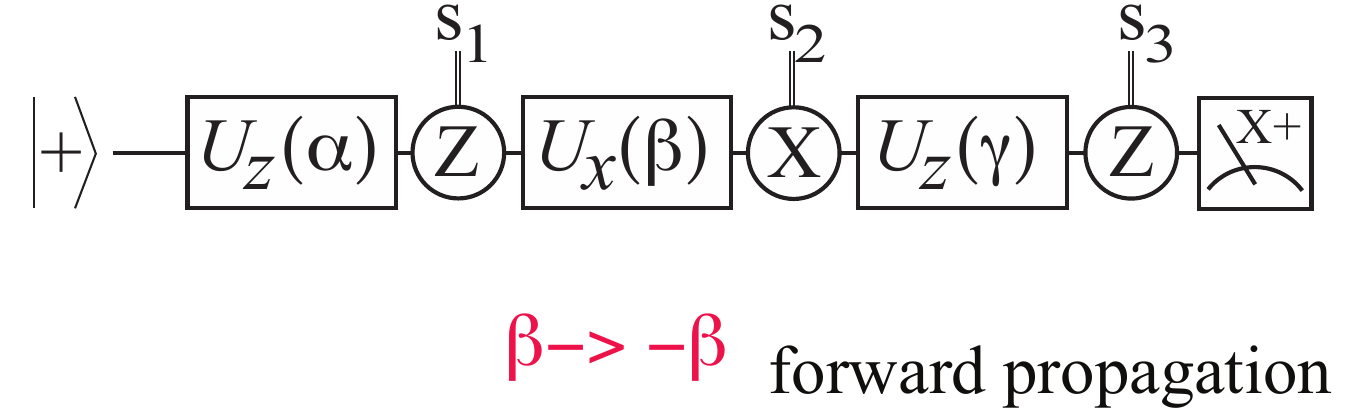} 
\end{tabular}
\end{center}
\caption{\label{LinExpl}Origin of the linear processing relations (by example). (a) MBQC on a 3-particle cluster state. It can be used to simulate the circuit shown in (b), but--if run straightforwardly--executes the probabilistic circuit ({c}). The need to compensate for the random measurement outcomes enforces a temporal order among the measurements. For explanation see text.}
\end{figure}

We briefly review how the linear processing relations in MBQC \cite{RB01} come about. As an example, we consider the 3-qubit cluster state shown in Fig.~\ref{LinExpl}a as resource. The measured local observables are $\cos 2\phi_i\, X_i + \sin 2\phi_i\, Y_i$. This MBQC can be used to simulate the circuit shown in Fig.~\ref{LinExpl}b, consisting of the application of a general one-qubit rotation (in its Euler decomposition) to a state $|+\rangle \sim |0\rangle + |1\rangle$, followed by a measurement of the Pauli observable $X$. However, if all three measurements are performed simultaneously, with measurement angles $\phi_1=\alpha$, $\phi_2=\beta$ and $\phi_3 = \gamma$, then, rather than the desired circuit of Fig.~\ref{LinExpl}b, the probabilistic circuit of Fig.~\ref{LinExpl}c is realized. It differs from the desired circuit by the insertion of Pauli spin or phase flips which are conditioned on measurement outcomes obtained. These random flips need to be removed from the circuit. This task  can be accomplished by measuring the three qubits in sequence and adjusting measurement bases on the go. 

Consider the phase flip $(Z)^{s_1}$ next to the $z$-rotation $U_z(\alpha)$ in Fig.~\ref{LinExpl}a. It can be propagated forward through the computation, past the readout measurement. Due to the anti-commutation relation $ZX=-XZ$, on its course the Pauli operator $Z$ flips the rotation angle $\beta$ and the outcome $o$ of the readout measurement. If qubit No.~1 is measured before qubit 2, the conditional flip of  the rotation angle beta can still be accommodated by a conditional flip of the measurement angle $\phi_2$, namely $\phi_2 = (-1)^{s_1} \beta$. The two other probabilistic insertions of Pauli flips propagate in a similar fashion. The net result is that the sign $(-1)^{q_2}$ of the measurement angle $\phi_2$ is given by $q_2=s_1$, and similarly $q_3=s_2$. Futhermore, the output $o$ is $o=s_1 \oplus s_3$. These relations are all mod 2 linear. This property is a consquence of the (anti)commutation relations of Pauli operators, and generalizes to the universal case \cite{RB01}.

\end{document}